\documentclass[%
prd,%
$onecolumn,%
twocolumn,%
nofootinbib,%
]{revtex4}%

\usepackage{graphicx}

\begin{document}

\title{Second-order corrections to the wave function at origin\\ in muonic hydrogen and pionium}
\author{Vladimir G. Ivanov}
\affiliation{Pulkovo Observatory, 196140, St.Petersburg, Russia}
\affiliation{D. I. Mendeleev Institute for Metrology (VNIIM), St. Petersburg 198005, Russia}
\author{Evgeny Yu. Korzinin}
\affiliation{D. I. Mendeleev Institute for Metrology (VNIIM), St. Petersburg 198005, Russia}
\author{Savely G. Karshenboim}
\email{sek@mpq.mpg.de}
\affiliation{D. I. Mendeleev Institute for Metrology (VNIIM), St. Petersburg 198005, Russia}
\affiliation{Max-Planck-Institut f\"ur Quantenoptik, 85748 Garching, Germany}

%
%


\begin{abstract}
Non-relativisitic second-order corrections to the wave function at
origin in muonic and exotic atoms are considered. The corrections
are due to the electronic vacuum polarization. Such corrections
are of interest due to various effective approaches, which take
into account QED and hadronic effects. The wave function at origin
plays a key role in the calculation of the pionium lifetime,
various finite nuclear size effects and the hyperfine splitting.
The results are obtained for the $1s$ and $2s$ states in pionic
and muonic hydrogen and deuterium and in pionium, a bound system
of $\pi^+$ and $\pi^-$. Applications to the hyperfine structure
and the Lamb shift in muonic hydrogen are also
considered.
\end{abstract}


\maketitle

\section{Introduction}

A number of atomic effects and, in particular, in exotic atoms
involve in the non-relativistic approximation various local
operators, proportional to the $\delta$-function in coordinate
space. The related contributions are proportional to the squared
value of the wave function at origin $\vert\Psi_{\rm
NR}(0)\vert^2$. Two examples of such operators are the operator of
interaction of the muon spin and the nuclear spin in a muonic atom
(that is responsible for the hyperfine structure) and the
$\pi^+\pi^-\pi^0\pi^0$ vertex operator (that is responsible for
the lifetime of the pionium atom).

That is a common feature of various non-relativistic
approximations and various effective non-relativistic approaches,
which are based on a separation of low-energy and high-energy
physics. Physics of atomic scale contributes to the
non-relativistic wave functions, while the higher energies and
momenta are responsible for various contact terms. That is very
much similar to the operator approach in theory of strong
interactions.

Meanwhile, there is an important difference between `conventional'
atoms and various exotic atoms in the calculation of $\vert\Psi_{\rm
NR}(0)\vert^2$. In conventional (electronic) atoms the
non-relativistic wave function is in most problems determined by its
pure Coulomb value $\vert\Psi_C(0)\vert^2$ and  most of the
corrections are either relativistic or have many-body origin. In
contrast, in muonic and pionic atoms, there is a specific class of
non-relativistic corrections, which can be still described by a
non-relativistic potential. The orbiting mass $m$ in such atoms is
much higher than the electron mass $m_e$ and in
particular\footnote{Technically, in the non-relativistic case that
is the reduced mass that enters the equations. It may be somewhat
below the muon mass $m_\mu$. The smallest values are in systems of
$\overline{\mu}\mu$, $\pi\mu$ and $\pi^+\pi^-$: $m_{\mu\mu}=0.5\,
m_\mu$, $m_{\pi\mu}=0.569\dots\,m_\mu$ and $m_{\pi\pi}=0.660\dots\,m_\mu$.}
\[
m \geq m_\mu \simeq 207 \;m_e\;.
\]

The characteristic momentum in such atoms $Z\alpha m c$ is higher
or comparable to $m_e c$, and thus the electronic vacuum
polarization produces a non-relativistic potential with the radius
of $\sim \hbar/(m_e c)$, which is somewhat larger than the atomic
Bohr radius $\sim \hbar/(Z\alpha m c)$.

\begin{figure}[ptbh]
\begin{center}
\includegraphics[height=5.5cm]{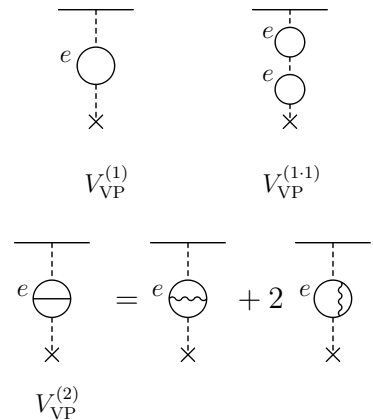}
\end{center}
\caption{Vacuum polarization corrections to electrostatic Coulomb
potential ($V_{\rm VP}$): the Uehling potential ($V^{(1)}_{\rm
VP}$), the reducible two-loop
 ($V^{(1\cdot1)}_{\rm VP}$) and irreducible  ($V^{(2)}_{\rm VP}$) two-loop potentials
  (the K\"allen-Sabry potential).\label{f:V}}
\end{figure}

The related potentials depicted in Fig.~\ref{f:V} modify the value
of the non-relativistic wave function at origin
\begin{eqnarray}
\vert\Psi_C(0)\vert^2\to \vert\Psi_{\rm
NR}(0)\!\!\!\!&\vert&\!\!\!\!^2 =
\vert\Psi_C(0)\vert^2\nonumber\\&\times&
\left(1+\frac{\alpha}{\pi}\,c_1+
\left(\frac{\alpha}{\pi}\right)^2\!c_2 +...\right)\!\!.
\end{eqnarray}

\begin{figure}[ptbh]
\begin{center}
 \includegraphics[height=2cm]{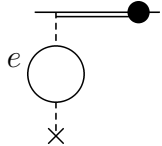}
\end{center}
\caption{The first-order correction to the wave function at
origin. The filled circle is for the $\delta({\bf r})$. The double
line stands for the non-relativistic reduced Coulomb Green
function.\label{f:1loop}}
\end{figure}

The related diagrams are presented in Fig.~\ref{f:1loop}. The
coefficient $c_1$ for the one-loop corrections is due to the Uehling
potential and was evaluated for a number of problems (see, e.g.,
\cite{muh_1s,pi_1s,lab,soto,muh_1s2s,2f1}).

\begin{figure}[ptbh]
\begin{center}
 \includegraphics[height=5.5cm]{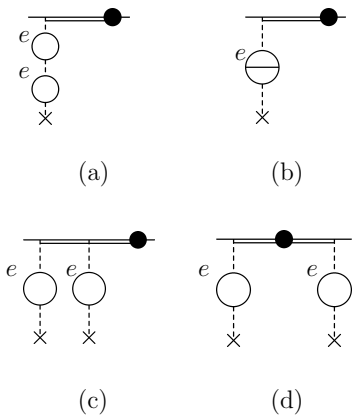}
\end{center}
\caption{The second-order vacuum polarization correction to the
wave function at origin.\label{f:2loop}}
\end{figure}

The second-order effects (see Fig.~\ref{f:2loop}) are due to
subsequent iterations of the Uehling term and due to
K\"allen-Sabry potential and here we present a calculation of
$c_2$ for the $1s$ and $2s$ states in muonic and pionic hydrogen
and deuterium and for the pionium atom.

\section{General expression for the correction to the wave function at origin: orders $\alpha$ and $\alpha^2$}

We consider a non-relativistic two-body system, which interaction
\begin{equation}
V(r) = V_C(r) + V_{\rm VP}(r)
\end{equation}
includes the Coulomb potential and its modification by the first-
and the second-order vacuum polarization (see Fig.~\ref{f:V})
\[
V_{\rm VP}(r)=V^{(1)}_{\rm VP}(r)+V^{(1\cdot1)}_{\rm
VP}(r)+V^{(2)}_{\rm VP}(r)\;.
\]
We are to find the wave function $\Psi_V(r)$ related to a
potential $V(r)$, or rather, only its value at origin. For the
latter, we can introduce an additional perturbation
\[
V_\delta (r) = A\,\delta({\bf r})
\]
and the obvious result of the perturbation theory, linear in $\delta
V_\delta (r)$, is
\[
\delta E_\delta = A\, \vert\Psi_{\rm NR}(0)\vert^2\;.
\]
That means that instead of finding the wave function we can
calculate the perturbation theory expansion for energy with the
perturbation $V_{\rm VP}(r)+V_\delta (r)$, taking terms linear in
$A$ and of a proper order in $\alpha$.

So, to find the coefficient $c_1$ it is enough to consider linear
terms in $V_{\rm VP}(r)$, and for the latter to apply the Uehling
term only. To find $c_2$ we have to evaluate two kinds of
contributions, which are
\begin{itemize}
\item[--] terms quadratic in $V_{\rm VP}(r)$, but including only the
Uehling potential ($V_{\rm VP}(r)\to V^{(1)}_{\rm VP}(r)$) (see
Fig.~\ref{f:2loop}$c$ and $d$);
 \item[--] terms linear in $V_{\rm VP}(r)$, which include the
second-order vacuum polarization (both reducible and irreducible:
($V_{\rm VP}(r)\to V^{(1\cdot1)}_{\rm VP}(r)+V^{(2)}_{\rm VP}(r)$)
(see Fig.~\ref{f:2loop}$a$ and $b$).
\end{itemize}

The results of the calculation of the coefficients $c_1$ and $c_2$
are collected in Table~\ref{t:cVP}.

 \begin{table}[ptbh]
 \caption{The results of the calculation of the coefficients $c_1$ and $c_2$ in various atoms for the $1s$ and $2s$ states. The $c_1$ coefficient was discussed in literature
 (see, e.g., \cite{muh_1s,pi_1s,lab,soto,muh_1s2s,2f1}), while the results for $c_2$ are found in
 this work. The accuracy of the presented results is determined by rounding of the numerical results.
 \medskip
 \label{t:cVP}}
 \begin{center}
 \begin{tabular}{|c|c|c|c|c|c|}
 \hline
 Atom & $m_r/m_e$ &\multicolumn{2}{|c|}{$1s$ } & \multicolumn{2}{|c|}{$2s$ }\\
 \cline{3-6}
  &  & $c_1$ & $c_2$& $c_1$ & $c_2$ \\
 \hline
 $\pi^-\pi^+$  &136.566& 1.35025 & 5.4378 & 1.13440 & 4.3723 \\
 $\mu H$       &185.841& 1.73115 & 7.2558 & 1.40425 & 5.5552 \\
 $\mu D$       &195.742& 1.80116 & 7.6038 & 1.45230 & 5.7730 \\
 $\pi H$       &237.764& 2.07748 & 9.0209 & 1.63850 & 6.6402 \\
 $\pi D$       &254.215& 2.17742 & 9.5504 & 1.70477 & 6.9584 \\
  \hline
 \end{tabular}
 \end{center}
 \end{table}

More details of the calculation of the $c_2$ coefficient can be
found in Table~\ref{t:c1sVP}, where we present separately all
contributions for the $1s$ state. The result for the
Fig.~\ref{f:2loop}$d$ contribution is split into two terms. That
reflects the fact that in general a contribution in the third
order of a perturbative theory is determined by the expression
(see, e.g., \cite{III,Bethe})
\begin{equation}\label{e3}
\Delta E^{(3)}(ns)=\langle \Psi_{ns}|\delta
 V\widetilde{G} \Bigl[\delta V-\Delta E_{ns}^{(1)}\Bigr]\widetilde{G}
 \delta V|\Psi_{ns}\rangle
 \;,
\end{equation}
where $\Delta E_{ns}^{(1)}=\langle\Psi_{ns}|\delta
V|\Psi_{ns}\rangle$, $\delta V$ is a sum of all perturbations
under consideration and $\Psi_{ns}
$ and $\widetilde{G}
$ are the wave function and the reduced Green function,
respectively, of the unperturbed problem (i.e., of the
non-relativistic Coulomb problem in our case).

A calculation of the subtraction term in (\ref{e3}), which is of
the form
\begin{equation}\label{sub:e3}
\Delta E^{\rm sub}(ns)=- \Delta E_{ns}^{(1)} \times \langle
  \Psi_{ns}|\delta
 V\widetilde{G}^2
  \delta V|\Psi_{ns}\rangle
  \end{equation}
is different from that of the main term in (\ref{e3}) and in fact
for the wave function it is even somewhat more complicated.

 \begin{table}[ptbh]
 \caption{Contributions
 to the value of $c_2(1s)$ for different diagrams in Fig.~\ref{f:2loop}. The result for the Fig.~\ref{f:2loop}$c$
 and $d$ contributions is split into two
 parts (cf. (\ref{e3})).
 \label{t:c1sVP}}
 \medskip
 \begin{center}
 \begin{tabular}{|c|c|c|c|c|c|}
 \hline
 Atom & $c^{\rm (a)}_2(1s)$ & $c^{\rm (b)}_2(1s)$ & $c^{\rm (c)}_2(1s)$& $c^{\rm (d)}_2(1s)$ & $c_2(1s)$ \\[1ex]
 \hline
  $\pi^-\pi^+$ & 1.3336 & 3.1031    & $0.6502-0.0719$ & $0.4558-0.0330$ & 5.4378    \\
 $\mu H$       & 1.8551 & 3.7967    & $1.0755-0.1525$ & $0.7492-0.0682$ & 7.2558    \\
 $\mu D$       & 1.9590 & 3.9166    & $1.1655-0.1719$ & $0.8110-0.0765$ & 7.6038    \\
 $\pi H$       & 2.3937 & 4.3693    & $1.5575-0.2635$ & $1.0790-0.1150$ & 9.0209    \\
 $\pi D$       & 2.5608  & 4.5253   & $1.7137-0.3032$ & $1.1853-0.1315$ & 9.5504    \\
  \hline
 \end{tabular}
 \end{center}
 \end{table}

To calculate the complete $\alpha^2$ corrections to any quantity one
indeed has to take into account relativistic corrections and
corrections to the operators, which are not universal. Let us briefly
discuss possible applications.

Let us consider specifically two systems, namely, the muonic
hydrogen and pionium.

(i) For the pionium a part of the $\alpha^2$ corrections is known
\cite{pi-report} and progress in experiment \cite{nemenov} requires
improvement of theory and calculation of remaining $\alpha^2$ terms,
which are in particular presented in this paper.

\begin{figure}[ptbh]
\begin{center}
\includegraphics[height=5.5cm,clip]{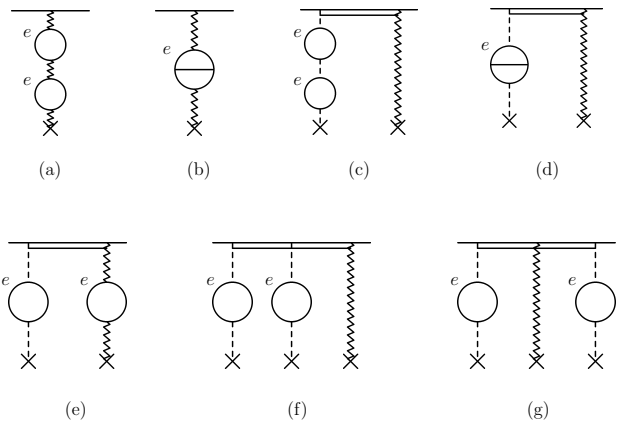}
\end{center}
\caption{The second-order vacuum polarization corrections to the
hyperfine splitting in muonic hydrogen. The wavy line stands for the
hyperfine interaction. \label{f:hfs-diag}}
\end{figure}

(ii) Another application of interest is due to the hyperfine
splitting in muonic hydrogen. In this case the complete result
includes various nuclear-structure dependent effects and the
$\alpha^2$ corrections are rather irrelevant for comparison with
expected experimental data both for the $1s$ \cite{muexp1s} and
the $2s$ \cite{muexp2s} hyperfine intervals. However, if both
experiments will deliver accurate results, one can consider a
specific difference \cite{muh_1s2s}
 \begin{equation}\label{d21def}
  \Delta E_{21}  = 8\times E_{\rm hfs}(2s) - E_{\rm hfs}(1s)\;,
 \end{equation}
and for this difference the calculation of the $\alpha^2$ terms
are relevant. The contribution induced by the correction to the
wave function is only a part of the complete result (cf.
Fig.~\ref{f:hfs-diag}) for the VP contributions, which is (see
\cite{russian} for detail)
 \begin{equation}
\Delta E^{VP}(1s)=\left(2.61419\,\frac{\alpha}{\pi} +
12.54584\left(\frac{\alpha}{\pi}\right)
 ^2\right)
\Delta E_{\rm hfs}^{(0)}(1s)\,,
 \end{equation}
  \begin{equation}
\Delta  E^{VP}(2s)=\left(2.31451\,\frac{\alpha}{\pi} +
10.65790\left(\frac{\alpha}{\pi}\right)
 ^2\right)
\Delta E_{\rm hfs}^{(0)}(2s)\,,
 \end{equation}
where
\begin{equation}
\Delta E_{\rm hfs}^{(0)}(ns)={8 \over 3}
  (1+a_\mu) \frac{\alpha(Z\alpha)^3 \,m c^2}{n^3} {m \over m_p}\, {\mu \over \mu_N}
\, \left({ m_{\rm r}\over m}\right)^3\;,
\end{equation}
$\mu_N$ is the nuclear magneton, $\mu$ stands for the proton
magnetic moment, and $m_{\rm r}$ is the reduced mass.

\begin{figure}[ptbh]
 \begin{center}
 \includegraphics[height=2cm]{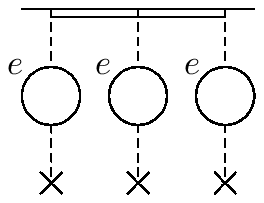}
 \end{center}
 \caption{The
$\alpha^5 m$ correction to the Lamb shift in muonic hydrogen: the
only contribution of the third order of non-relativistic
perturbation theory (see (\ref{e3})).
 \label{f:Lamb3}}
 \end{figure}

Performing a calculation of third terms depicted in
Fig.~\ref{f:2loop}$d$ and Fig.~\ref{f:hfs-diag}$g$ as a test, we
also calculated a contribution to the Lamb shift (see
Fig.~\ref{f:Lamb3}), which was previously calculated for muonic
hydrogen in \cite{kinoshita}.

The result \cite{kinoshita} was for the correction to the Lamb
shift (i.e. for a splitting of the $2s$ and $2p$ states). This
value is of a particular interest because of the PSI experiment
\cite{muexp2s}. We have also  calculated the same quantity and
found
\[
\Delta E^{\rm
(Fig.~\protect\ref{f:Lamb3})}_{2p-2s}(\mbox{this~work})
=
\Biggl[\bigl(-7.3861\cdot10^{-6}+0.3511\cdot10^{-6}\bigr)
\]
\begin{equation}\label{ours}
- \bigl(-0.002\,5412+0.001\,3661\bigr)\Biggr]
\frac{\alpha^5}{\pi^3} m_r c^2
= 0.0011681\frac{\alpha^5}{\pi^3} m_r c^2\;,
\end{equation}
where the first parenthesis is for the $2p$ contribution, while
the second is for the $2s$ one; each parenthesis consists of a
main term and a subtraction term as introduced in (\ref{e3}).

Our result disagrees with the result published in \cite{kinoshita},
\begin{equation}\label{kn}
\Delta E^{\rm
(Fig.~\protect\ref{f:Lamb3})}_{2p-2s}(\mbox{Ref.~\protect\cite{kinoshita}})=0.002535(1)\frac{\alpha^5}{\pi^3}
 m_r c^2\;.
\end{equation}

After the work was finished we contacted the authors of
\cite{kinoshita}. As the results of the communications it has been
agreed that our result for the main terms for both states ($2s$ and
$2p$) confirm calculations in \cite{kinoshita}, while the
subtraction term was missing there. After their result was corrected
for subtraction, it agrees with ours (see our eprint \cite{eprint}
for detail.


Concluding, we calculated non-relativistic corrections in the
relative order $\alpha^2$ to the wave function at origin in muonic
and exotic atoms for the $1s$ and $2s$ states, presented a result
on the non-relativistic $\alpha^2$ correction to the hyperfine
structure in muonic hydrogen for the same states and corrected the
result \cite{kinoshita} on the $\alpha^5m_\mu c^2$ contribution to
Lamb shift in muonic hydrogen. Details of our calculations of the
second-order vacuum polarization effects for the wave function at
origin in various atoms and the hyperfine splitting and the Lamb
shift in muonic hydrogen are in preparation and will be published
elsewhere.

\section*{Acknowledgment}

The work was in part supported by DFG (under grant \# GZ 436 RUS
113/769/0-3) and RFBR (under grant \# 08-02-91969). The authors
are grateful to V. Lyubovitskij, A. Rusetsky, and V. Shelyuto for
useful and stimulating discussions. A part of work was performed
during visits of VGI and EYK to Max-Planck-Institut f\"{u}r
Quantenoptik and they are grateful for its hospitality. Work of
EYK was also supported by the Dynasty foundation.
The authors are grateful to T. Kinoshita and M. Nio for communication
on details of their calculations.

\end{document}